\documentstyle[aps,epsf,pre,multicol]{revtex}
\draft

\begin{document}
\title{Boundary Limitation of Wavenumbers in Taylor-Vortex Flow}
\author{Marcus Linek$^{(1)}$ and Guenter Ahlers$^{(2)}$}
\address{$^{(1)}$Fachbereich Produktionstechnik}
\address{Universit\"at Bremen}
\address{D28359 Bremen, Germany}
\address{$^{(2)}$Department of Physics and Center for Nonlinear Science}
\address{University of California at Santa Barbara}
\address{Santa Barbara, CA 93106.}
\date{\today}

\maketitle

\begin{abstract}

\noindent

We report experimental results for a boundary-mediated wavenumber-adjustment mechanism and for a boundary-limited wavenumber-band of Taylor-vortex flow (TVF). The system consists of fluid contained between two concentric cylinders with the inner one rotating at an angular frequency $\Omega$. As observed previously, the Eckhaus instability (a bulk instability) is observed and limits the stable wavenumber band when the system is terminated axially by two rigid, non-rotating plates. The band width is then of order $\epsilon^{1/2}$ at small $\epsilon$ ($\epsilon \equiv \Omega/\Omega_c - 1$) and agrees well with calculations based on the equations of motion over a wide $\epsilon$-range. When the cylinder axis is vertical and the upper liquid surface is free (i.e. an air-liquid interface), vortices can be generated or expelled at the free surface because there the phase of the structure is only weakly pinned. The  band of wavenumbers over which Taylor-vortex flow exists is then more narrow than the stable band  limited by the Eckhaus instability. At small $\epsilon$ the boundary-mediated band-width is linear in $\epsilon$. These results are qualitatively consistent with theoretical predictions, but to our knowledge a quantitative calculation for TVF with a free surface does not exist. 

\end{abstract}

\begin{multicols}{2}

\section{Introduction}
\label{intro}

Many physical systems will undergo a transition from a spatially-uniform state to
a state with spatial variation when they are driven sufficiently far from equilibrium by
an external stress. This spatial variation is known as a "pattern".
Typical of such systems are Rayleigh-Ben\'{a}rd convection (RBC) \cite{RBC} and
Taylor-vortex flow (TVF) \cite{TVF}. The present paper will focus on TVF, which is a flow pattern of a fluid confined between two concentric cylinders.  When rotating the inner cylinder about its axis, the initial instability which occurs as the angular frequency $\Omega$ is increased is a transition at $\Omega_c$ from a uniform (circular Couette) state to the TVF state. TVF is periodic in the axial direction and has full rotational symmetry in the azimuthal direction. In the axial direction the pattern is characterized by a wavenumber $k = 2\pi/\lambda$, where $\lambda$ is the wavelength. One wavelength consists of a vortex pair since adjacent vortices have opposing circulation.  
For $\Omega > \Omega_c$, TVF solutions to the equations of motion exist over a band of wavenumbers which depends on $\epsilon \equiv \Omega/\Omega_c - 1$ and on the boundary conditions in the axial direction.

We refer to patterns with non-trivial variation only in one direction (such as TVF) as one-dimensional patterns. It is well known that the band of wavenumbers of one-dimensional patterns of finite length, with the pattern phase pinned at each end, is limited by the marginal curve where perturbations of the uniform state first acquire a positive growth rate. In TVF with rigid, non-rotating ends a large-amplitude Ekman vortex forms adjacent to each end and is associated with phase pinning. However, the resulting solutions of such a system are not all stable. The (more narrow) stable band of states is limited by the long-wavelength Eckhaus instability.\cite{Ec65,CDA83,KZ85,ZK85,DCA86,RP86,HAC86,ACDH86,PK87,KSZ88p,PR88bp,DCA94} The Eckhaus instability is a bulk instability which manifests itself in the system interior where one pattern wavelength (one pair of Taylor vortices) is either gained or lost, depending on whether $k$ is larger or smaller than the stable band limited by $k_E(\epsilon)$. One of the successes in the study of pattern-forming systems is the agreement between the calculated and measured $k_E(\epsilon)$ for TVF which has been found with three different ratios of the cylinder radii.\cite{CDA83,DCA86,HAC86,ACDH86,DCA94}

Theoretically, phase pinning (and thus the Eckhaus instability) at small $\epsilon$  is expected when the boundary conditions at the system ends $z = 0, L$ correspond to a large amplitude $A(0) = A(L) = A_0$ of the velocity field, say $A_0 = {\cal O}(1)$, while in the system interior the amplitude $A(z)$ is small, say ${\cal O}(\epsilon^{1/2})$. This situation closely corresponds to the TVF system with rigid ends where the influence of the Ekman vortex can be approximated by this boundary condition.\cite{GD82,DCA86,ACDH86} The Eckhaus-stable band then has the same width as the band of stable states for the infinite system,\cite{finite-size} but the number of states is finite since only discrete states with an integer number of vortices $N = 2 L/\lambda = L k/\pi$ can occur.

It was shown theoretically by Cross et al.\cite{CH92,CDHS} (CDHS) and discussed from various viewpoints by others\cite{PZ80,Po81,KH83,Po83,HKR85,KZ85,ZK85,PK87} that a reduction of $A_0$ to  $A_0 = \tilde \lambda \epsilon^{1/2}$ with $\tilde \lambda \alt {\cal O}(1)$ leads to a qualitative change of the width of the wavenumber band over which solutions exist. Solutions corresponding to such boundary conditions are known as type I solutions.\cite{KH83} The limit of their existence is determined by  phase slip at the boundary. The solutions appear to be stable over their entire existence range. The band limits $k_{b}^+$ and $k_{b}^-$ at large and small $k$ respectively vary linearly with $\epsilon$ when $\epsilon$ is small, yielding  $k_{b}^- = \lambda^- \epsilon$ and $k_{b}^+ = \lambda^+ \epsilon$ where $\lambda^{\pm}$ are constants which depend on the particular system. Thus a dramatic reduction of the band at small $\epsilon$ from ${\cal O}(\epsilon^{1/2})$ (in the Eckhaus case) to ${\cal O}(\epsilon)$ is predicted. Solutions corresponding to $\tilde \lambda > {\cal O}(1)$ are called type II solutions;\cite{KH83} they exist over a wide range inside the marginal-stability curve and are stable over the entire Eckhaus-stable band. 

The reduction of the band and the occurrence of phase slip near the sidewall was verified by experiments on RBC\cite{MHLPP82,MHLPP84}, on the buckling of plates subjected to a compressional stress\cite{WB83,BWG84}, and on B\'enard-Marangoni convection\cite{Be88}. However, this early work was mostly qualitative and for relatively large $\epsilon$. Qualitative confirmation of a boundary-reduced wavenumber band was obtained also from numerical calculations for narrow RBC systems of finite length.\cite{MHL86,ABN87} 

More recently, Mao et al. investigated boundary-limitation of the wavenumber-band in a long, narrow, and very thin (two-dimensional) film of a smectic-A liquid crystal which undergoes convection when driven by an electric field.\cite{MDDM96} The convection rolls form a one-dimensional pattern. Measurements showed that the flow velocity of the rolls is strongly reduced near the ends of the system, suggesting that the boundary conditions correspond to a value of $\tilde \lambda$ much less than one. Thus one would expect the boundary-mediated mechanism to come into play. Indeed it was observed in the experiment that convection rolls were always lost or gained at the system ends and never in the interior. For this system the location of the Eckhaus stability limit was estimated from the location of the neutral curve, which in turn was calculated from the equations of  motion\cite{DMB98}. Thus a comparison of the reduced boundary-limited band with the approximate location of the wider Eckhaus band of the infinite system is possible. The results include data over the range $0 < \epsilon < 1$. They convinvingly demonstrate the reduced band width and are consistent with a linear dependence of $k_b^\pm - k_c$ on $\epsilon$ in the range $\epsilon < 1$ ($k_c$ is the critical wavenumber which forms for $\epsilon = 0$). 

Here we report experimental results for the wavenumber band of TVF. Using rigid, non-rotating ends, we reproduced the well known Eckhaus mechanism and stability range.\cite{DCA86,RP86}. We then investigated a system with a free liquid-air interface which terminated the fluid axially in an apparatus oriented vertically with its axis. Since the influence of the viscosity of air and surface-tension effects presumably were small, one would expect such a surface to be approximated reasonably well by free-slip boundary conditions in the theory, and these would lead to an amplitude $A_0$ equal to that in the bulk (i.e. $\tilde \lambda \simeq 1$). Semi-quantitative verification of this was obtained from the flow visualizations to be presented below. Thus, we expect this TVF case to be on the borderline where it would be difficult to say a priori whether the observable band is limited by the Eckhaus (bulk) instability or whether the existing band is limited by phase slip at the boundary.
We demonstrated experimentally that the free liquid-air interface provides sufficiently weak phase pinning to permit phase slip. Thus vortices were gained or lost at that surface rather than in the bulk. This mechanism led to a narrower band than the Eckhaus-stable band, with boundaries at $k_b^\pm(\epsilon)$ which at small $\epsilon$ were linear in $\epsilon$ as expected theoretically. However, these boundaries did not pass through $k_c$ and $\epsilon = 0$. Instead there remained a small gap. It is not clear whether this gap was due to the finite length of our system\cite{finite-size} or whether it was associated with the marginal boundary conditions ($\tilde \lambda \simeq 1$).

Although it is peripheral to the main purpose of this paper, we note that the radial velocity component at the free surface was almost always found to be in the outward direction. This is consistent with free-slip boundary conditions, where the centrifugal force is unopposed. It differs from the rigid boundary near which normally the flow is inward because the rigid boundary condition of vanishing velocity can not sustain the radial pressure gradient required by outward flow. Thus, with one rigid and one free boundary, our system normally contained an odd number $N$ of vortices. However, anomalous states with inflow at the free surface (and thus an even $N$) were also encountered on rare occasions.   

In Sect.~\ref{expt} of this paper we describe the experimental apparatus and procedures,  and the methods of analysis which were used. The experimental
results for the instability mechanism and the stability band are presented in Sect.~\ref{results}. A brief discussion of the results and some ideas for future work follow in Sect.~\ref{discussion}.

\section{Apparatus and Procedures}
\label{expt}

\begin{figure}  
\epsfxsize = 3in  
\centerline{\epsffile{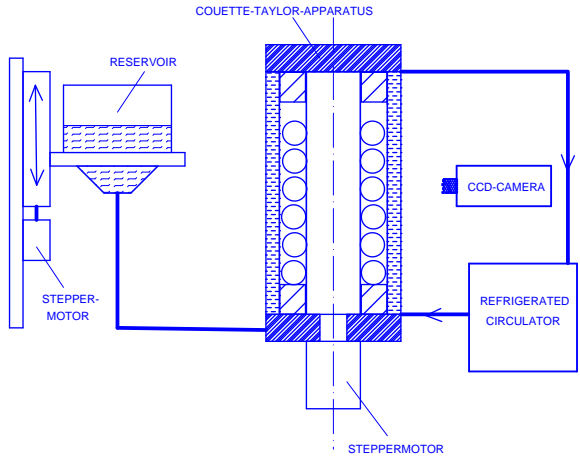}}  
\vskip 0.2in
\caption{Schematic diagram of the apparatus.}  
\label{fig:apparatus}
\end{figure}

A schematic diagram of the apparatus is shown in Fig.~\ref{fig:apparatus}. The Couette-Taylor column was similar to the one described by Dominguez-Lerma et al. \cite{DCA86}. It consisted of two concentric, straight cylinders. 
The outer one was made of plexiglass. The inner cylinder had an aluminum core which was clad with delrin. The radii were $r_1 = 18.68$ and $r_2 = 25.44$ mm, yielding a radius ratio
$\eta \equiv r_1/r_2 = 0.734$ and a radial gap $d \equiv r_2 - r_1 = 6.76~mm$.
A computer-controlled stepper motor rotated the inner cylinder with an angular frequency $\Omega = 2\pi f$, and the
outer cylinder was at rest. 
The system was temperature controlled to about $\pm$ 20 mK by water from a Neslab refrigerated circulator. This water flowed through a transparent jacket surrounding the outer cylinder.

\narrowtext

The gap between the cylinders contained two non-rotating delrin collars, one located near each end of the apparatus. The collars had a flat surface facing the fluid and orthogonal to the cylinder axis. The axial position of the upper collar was adjustable by means of two $2.5 \ {\rm mm}$ diameter stainless-steel rods which passed through seals in the upper end cap of the apparatus and attached to the collar. We were able to create either a rigid or a free
(liquid/air) surface  at the top of the column by adjusting the liquid level and/or the top collar position. This also allowed for changes of the aspect ratio $L \equiv H/d$, where $H$ is the distance in mm between the rigid boundaries or between the bottom boundary and the free surface. Typically, we used $L \simeq 40$ to 50.

The fluid level
in the Taylor column could be adjusted by changing the 
vertical position of the reservoir shown in Fig.~\ref{fig:apparatus}. This reservoir was attached to a vertical translation stage which could be moved with a second computer-controlled stepper motor.  

For the working fluid we used a mixture of 50\% glycerol, 
48\% water, and 2\% Kalliroscope for flow
visualization. This mixture had a kinematic viscosity of about 0.07 cm$^2$/s, yielding a gap-diffusion time $t_\nu \equiv d^2/\nu \simeq 7$ sec. The axial diffusion time $\L^2t_\nu$ then is about 4 hours for the typical values $L \simeq 45$ used by us. Typical axial equilibration times required to reach an axially uniform state are determined by phase diffusion and depend on the wavenumber. For small $\epsilon$ and $k$ near $k_c$ they are expected\cite{Sn69,Ah89} to be  about $0.2 L^2 t_\nu \simeq 50$ minutes. Near the Eckhaus boundary they are expected to become much longer because the phase diffusivity vanishes at that boundary. We depended upon the measured time evolution of the phase of our system to decide whether the data were quasi-static. Amplitude equilibration after a change of $f$ is generally much faster and typically occurs on a time scale of $\tau_0 t_\nu / \epsilon$ with\cite{DAC84} $\tau_0 = 0.076$, which for $\epsilon \agt 10^{-2}$ is only about a minute. 

\begin{figure}
\epsfxsize = 3in  
\centerline{\epsffile{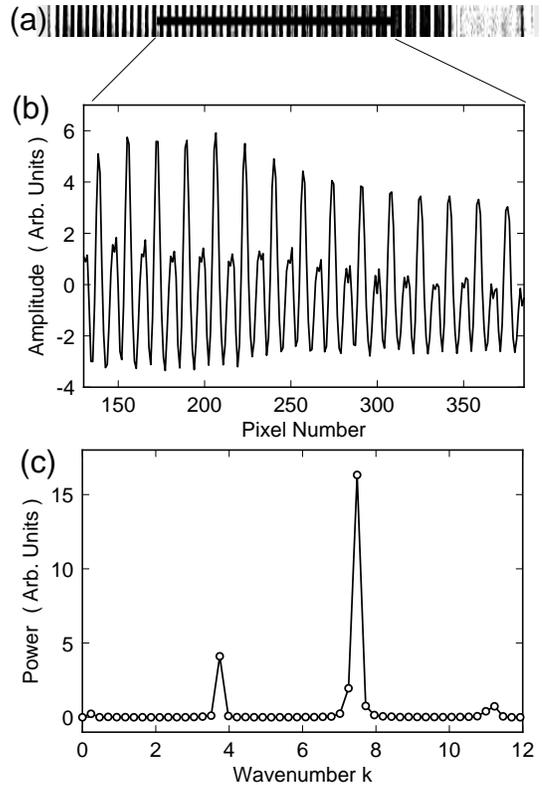}}  
\caption{Illustration of the image analysis. The top is an example of a single image (a ``snapshot"). The middle is a plot of the image amplitude, averaged in the narrow (azimuthal) direction, as a function of the axial direction in units of camera pixels. The bottom is the Fourier transform of the amplitude (after length was scaled by the gap $d$) and shows the wavenumbers of the Taylor vortices.}  
\label{fig:analysis}
\end{figure}  

A CCD camera was used to digitally record an image of the Taylor-vortex pattern.
Typically, we took images with a width (in the azimuthal direction) of 30 pixels and a length (in the axial direction) of 534 pixels. An example is shown in Fig.~\ref{fig:analysis}a. Division by a reference image taken below the onset of TVF  was used to reduce inhomogeneities due to uneven illumination. The pixels were averaged over the width to produce a one-dimensional record of the TVF state as a function of the axial position. Temporal sequences of such one-dimensional records were used to prepare a two-dimensional space-time ``image" which showed the evolution of the TVF state as a function of time and/or inner-cylinder frequency (see for instance Figs.~\ref{fig:onset}, \ref{fig:mechanism}, and \ref{fig:detail} below). In order to avoid effects associated with the cylinder ends (such as the strong Ekman vortex at the rigid bottom boundary), only a central segment of length 256 pixels of each one-dimensional record was used for further analysis. The section used is shown in Fig.~\ref{fig:analysis}a by the black bar. An example is plotted in Fig.~\ref{fig:analysis}b. The variation of the signal amplitude with axial position is due primarily to effects associated with uneven illumination which remained even after image division.  

\begin{figure}  
\epsfxsize = 2.5in  
\centerline{\epsffile{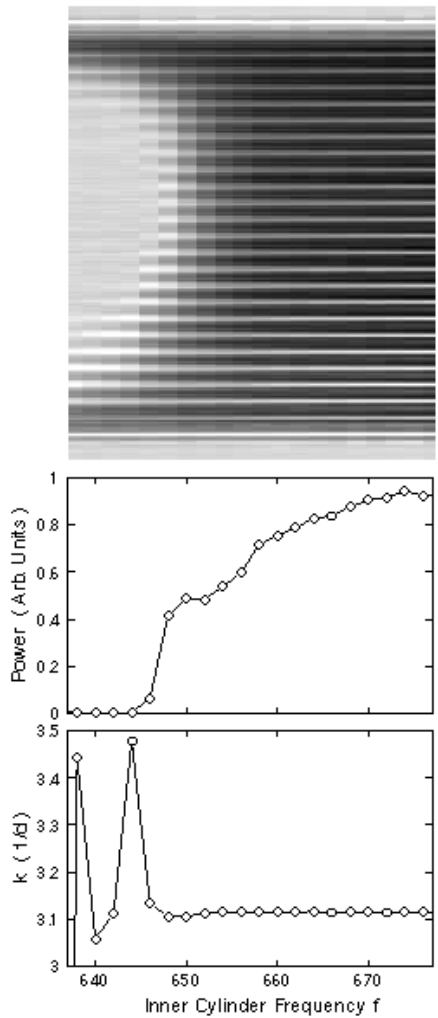}}  
\vskip 0.2in
\caption{An illustration of the determination of $\Omega_c$. The top image is a space-frequency (or time) plot with an expanded frequency scale near onset. The middle gives the signal power determined from the Fourier transform (see Fig.~\protect{\ref{fig:analysis})}. The bottom shows the corresponding wavevector determinations.}  
\label{fig:onset}
\end{figure}

The data were multiplied by a Hanning window function and then Fourier transformed. The square of the modulus of such a transform is shown in Fig. ~\ref{fig:analysis}c as a function of the wavenumber $k$. It is apparent that the second harmonic is significantly stronger than the fundamental. This is expected from the Kalliroscope flow visualization since outflow and inflow boundaries of vortices have a similar appearance. We computed the first moment of the second harmonic, using typically only the three points nearest the maximum, to determine the wavenumber of the TVF structure. To determine the onset frequency $f_c$, we also determined the power under the second-harmonic peak. 

In order to determine $f_c$, a TVF state was usually prepared first by slowly increasing the cylinder frequency $f$ to a value above the critical value $f_c$, and by waiting for equilibration of this state. Then $f$ was stepped down in increments of 2 mHz, with an equilibration time of 100 to 600 seconds after each step. On average this corresponds to a ramp rate $\beta \equiv (t_\nu/f_c)(df/dt) \alt 10^{-4}$ (here $t$ is in sec). The image in the top of Fig.~\ref{fig:onset} shows the evolution of the axially-varying amplitude for the section of such a run very near $f_c$ as a function of time and thus of frequency. Here the individual time/frequency steps are resolved. For this case there was a rigid boundary at each end. The Ekman vortices and their influence on the internal TVF structure are apparent. However, the system is sufficiently long\cite{DCA86} that the ends do not significantly influence the determination of $f_c$ in the interior. The middle part of Fig.~\ref{fig:onset} gives the power of the second harmonic of the Fourier transform (determined as discussed above) as a function of $f$ over the same range as the image above it. For the example in the figure it is easy to see that $f_c = 646 \pm 2$ mHz. Increasing the frequency from below $f_c$ gave the same results within the quoted uncertainties.
The bottom part of Fig.~\ref{fig:onset} shows the wavenumber derived from the same run. One can see that $k$ is determined unambiguously by the data at all frequencies greater than $f_c$.

\begin{figure}
\epsfxsize = 3in  
\centerline{\epsffile{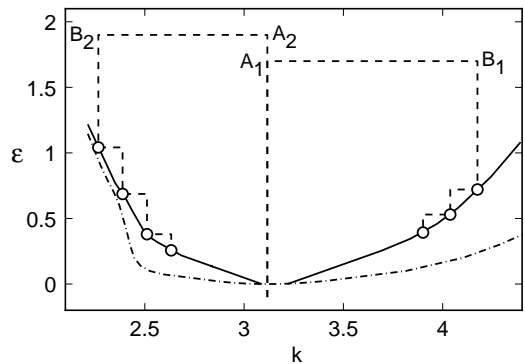}}  
\caption{Illustration of the experimental protocol (see text for details). The solid line is a smooth curve passing through our results for the boundary-mediated stability limit. The dash-dotted curve is the Eckhaus stability limit.}  
\label{fig:protocol}
\end{figure}  

Two examples of an experimental protocol designed to determine the boundary-selected stability limits are illustrated in Fig.~\ref{fig:protocol}. When the large- (small-) $k$ stability boundary was to be investigated, a liquid column with a free surface and with a large (small) $L = L_0$ was first prepared. The examples in Fig.~\ref{fig:protocol} correspond to $L_0 = $61 (37). An initial state with $k$ near $k_c$ was then prepared by slowly ramping $f$ from below to above $f_c$. A state with  $N = $ 61 (37) vortices resulted. Next the frequency was increased relatively quickly to two to four times $f_c$ (points A$_1$ and A$_2$ in the figure), generally without changing the number of vortices. The wavenumber of the system was then adjusted to a desired starting value [point B$_1$ (B$_2$) in the figure] by lowering (raising) the liquid level in the TVF column. This was accomplished by decreasing (increasing) the vertical reservoir position. When the reservoir position was changed sufficiently slowly, no loss or gain of vortices occurred provided the structure remained well within the stable wavenumber band. Typically we changed the reservoir position at the rate $dz/dt \simeq 10^{-3}$ cm/s, corresponding to $(t_\nu/d)(dz/dt) \simeq 10^{-2}$. Typically one or two hours were required to reach points B$_1$ or B$_2$. In the examples of Fig.~\ref{fig:protocol} $B_1$ ($B_2$) corresponded to $L =$ 45.9 (51.3).

Having established states B$_1$ or B$_2$, the frequency was reduced in steps of 2 mHz at time intervals of typically 100 to 600 seconds, corresponding to effective ramp rates near $\beta = -10^{-4}$. Before each new step, an image was taken. The circles in Fig.~\ref{fig:protocol} illustrate idealized sequences of instability points which are encountered in this process. Actual experimental runs will be illustrated in the next Section by space-time images and by plots of the wavenumber as a function of time or frequency.

The procedure for determining the Eckhaus boundary was similar, except that at the state indicated by B$_1$ and B$_2$ in Fig.~\ref{fig:protocol} the rigid collar bounded the fluid at the top. Sudden-start protocols\cite{BK74} were used as well at times to quickly prepare initial states with $k$ different from $k_c$.     

\section{Results}
\label{results}

\begin{figure}  
\epsfxsize = 3.375in  
\centerline{\epsffile{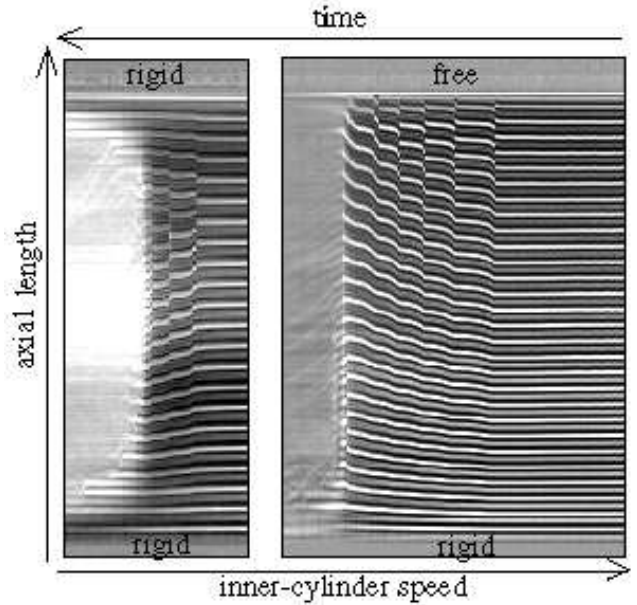}}  
\vskip 0.2in
\caption{Illustration of the two mechanisms for wavenumber limitation.}  
\label{fig:mechanism}
\end{figure}

Figure~\ref{fig:mechanism} gives space-time images which illustrate the two different mechanisms. The left part is a run with rigid top and bottom ends, and with $k < k_c$. The Ekman vortices and their influence on the amplitude of the Taylor vortices in the interior but near the two ends are apparent. The instability mechanism led to the gain of vortex pairs in the system interior. Three transitions are clearly visible, although the third was almost immediately followed in the sample interior by the axially-uniform Couette state.

A similar experiment with a free surface yielded a very different result. This is illustrated in the right part of the figure. In this case the original state (near the right of the image) has $k > k_c$. Five or more transitions can be seen, but vortex pairs were lost at the free top surface rather than in the interior.

\begin{figure}  
\epsfxsize = 3in  
\centerline{\epsffile{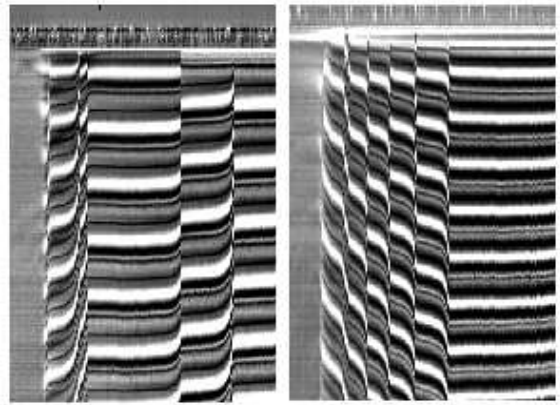}}  
\vskip 0.2in
\caption{Detailed view of the losses (right image) and gains (left image) of vortices at the free upper surface.}  
\label{fig:detail}
\end{figure}

The phase slip at the free surface is illustrated in greater detail in Fig.~\ref{fig:detail}. The left example is for $k < k_c$. In this case vortices were added at the free surface, so as to increase $k$. The right one is for $k > k_c$ where vortices were expelled so as to decrease $k$. Figure~\ref{fig:detail} also illustrates the nature of the boundary condition at the free surface. The TVF amplitude is essentially constant as a function of $z$, with at most a very small enhancement (depression) visible for $k < k_c$ ($k > k_c$) as the Couette state is entered. The very small difference, if any, between the two cases shown probably is due to varying conditions with height of the cylinder surfaces rather than a dependence on $k$. 

\begin{figure}  
\epsfxsize = 3in  
\centerline{\epsffile{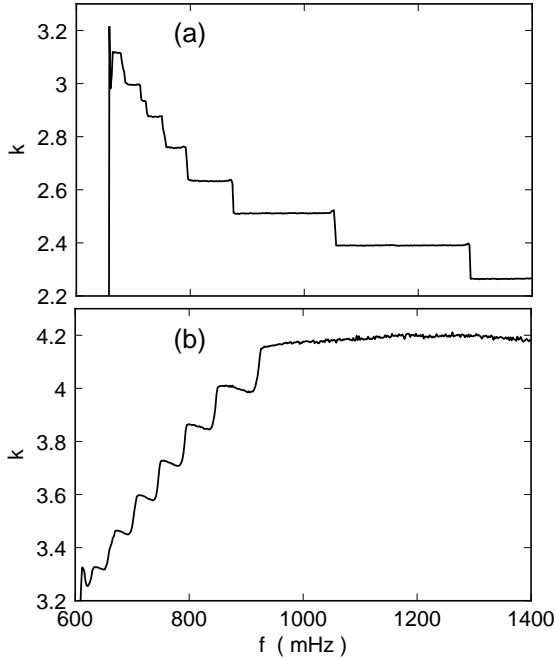}}  
\vskip 0.2in
\caption{The wavenumber $k$ as a function of the inner-cylinder frequency $f$ . Temporally, the experiment proceeded from right (large $f$) to left (small $f$). (a): $k < k_c$. (b): $k > k_c$.}  
\label{fig:k}
\end{figure}

Figure~\ref{fig:k}a gives the wavenumber as a function of $f$ for a run which started with $k < k_c$.  The parameters are those used in Fig.~\ref{fig:protocol} to illustrate the small-$k$ experiments. Here $L = 51.3$, and initially (at the right of Fig.~\ref{fig:k}a) there were $N = 37$ vortices. A gain of a vortex pair corresponds to $\delta N = 2$ and should yield $\delta k = (\pi /L) \delta N = 0.122$ if the anomalous width of the Ekman vortex at the bottom boundary is neglected. The average measured step size was also $\delta k = 0.122$, consistent with the expected effect of a pair gain. It is clear from Fig.~\ref{fig:k}a that most of the transitions are pair gains, thus maintaining an odd number of vortices in the system. This is consistent with the preference for outflow at the free and inflow at the rigid surface. However, there are occasional exceptions. One of them can be seen in Fig.~\ref{fig:k}a near $f = 722$ mHz, where two successive single-vortex-gains ($N = $ 47 to 48 to 49) resulted in smaller values  $\delta k \simeq 0.058$.  

It can be seen that the system equilibrates surprisingly fast after each transition. This is qualitatively consistent with calculations of the phase diffusivity $D(k,\epsilon)$, which show a strong asymmetry as a function of $k$ at constant $\epsilon$  with a maximum at relatively small $k < k_c$.\cite{Ri86} Equilibration is indeed much slower for $k > k_c$, as can be seen in Fig.~\ref{fig:k}b. For this case $L = 45.9$ and $N = 61$ at the beginning of the run. After each transition there is a noticeable slow relaxation of the average wavenumber in the section of the system used for the analysis. Nonetheless, reasonable estimates of the stability boundary can be obtained from the experiment. In this case the steps have an average value $\delta k = 0.141$. For pair losses and this $L$ one would expect 0.136, in good agreement with the experiment.  

\begin{figure}  
\epsfxsize = 3in  
\centerline{\epsffile{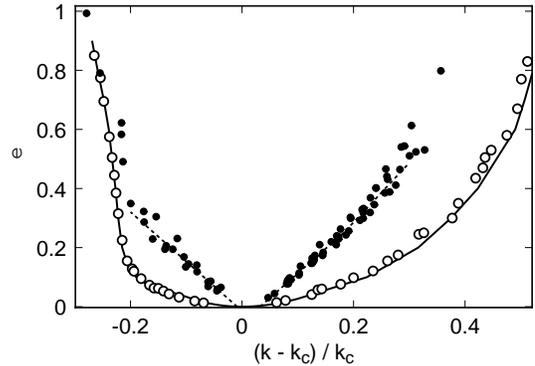}}  
\vskip 0.2in
\caption{Results for the wavenumber-band limits in the presence of a free boundary (solid circles). Also shown (open circles) are the results from Ref. \protect{\cite{DCA86}} for the Eckhaus boundary and the theoretical Eckhaus boundary (solid line) from Ref.\protect{\cite{RP86}}.}  
\label{fig:results}
\end{figure}

Figure~\ref{fig:results} summarizes all our determinations of the stability boundary with a free surface as solid circles. They were obtained from numerous runs with different values of $L$ over a time period of about 4 months, each run giving a few points on one of the boundaries. They scatter somewhat more than we anticipated on the basis of our resolution for $\epsilon$ and $k_s^\pm$. We believe that there may be a contribution to the scatter from irreproducibilities from run to run in the nature of the air-liquid interface. This interface could for instance be influenced by contaminations of the cylinder surfaces by the Kalliroscope, and this contamination could vary with $L$; but we do not have any direct information about this. For reference, we show the theoretical Eckhaus boundary as a solid line and the experimental determinations of the Eckhaus boundary by Dominguez-Lerma et al. as open circles. Regardless of their scatter, it can be seen that the free-surface-limited data 
lie well inside the Eckhaus-stable band over the entire $\epsilon$-range of the experiment.

\begin{figure}  
\epsfxsize = 3in  
\centerline{\epsffile{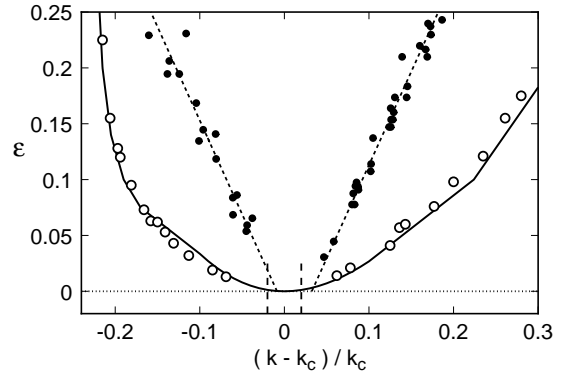}}  
\vskip 0.2in
\caption{Results at small $\epsilon$ for the wavenumber-band limits in the presence of a free boundary (solid circles). Also shown (open circles) are the results from Ref. \protect{\cite{DCA86}} for the Eckhaus boundary and the theoretical Eckhaus boundary (solid line) from Ref.\protect{\cite{RP86}}.}  
\label{fig:results_detail}
\end{figure}

 At small $\epsilon$ the data are consistent with a linear dependence on $\epsilon$, as shown by the dashed lines in the figure. The small-$\epsilon$ portion of the results is shown once more on expanded scales in Fig.~\ref{fig:results_detail}. The linear dependence of $k_s$ upon $\epsilon$ is clearly seen. The slopes $k_cd\epsilon/dk_b^\pm$ of the boundaries are within error symmetric and equal to about $\pm 1.7$. However, the straight lines through the data do not pass through the origin. Instead, there is a small gap near $k_c$ which reaches approximately from $\delta k^-/k_c = -0.01$ to $\delta k^+/k_c = 0.03$. Whereas the width $\delta k^+/k_c - \delta k^-/k_c \simeq 0.04$ of the gap has an uncertainty of only about 0.01, the values of $\delta k^+/k_c$ and $\delta k^-/k_c$ themselves are less well known because of possible small systematic errors in $k$. Thus a gap symmetric about $k_c$ is not ruled out.

\section{Discussion}
\label{discussion}

We presented experimental results which demonstrate that TVF terminated at one end by a liquid-air interface has a band of solutions which is more narrow than the Eckhaus-stable band of the system with rigid non-rotating ends. The band width is limited by phase slip at the free surface. At small $\epsilon$, the band limits $k_b^\pm(\epsilon)$ vary linearly with $\epsilon$. This is in agreement with general theoretical predictions, but to our knowledge specific calculations for TVF have not been carried out. The boundaries $k_b^\pm(\epsilon)$ do not pass through $k_c$ at $\epsilon = 0$ and instead leave a small wavenumber gap $\delta k^+ - \delta k^-$ at the neutral curve which was not theoretically predicted. 

At this point we can only speculate about the origin of the gap. Perhaps it is asociated with the particular boundary conditions which pertain to our physical system. As discussed above, we expect TVF with a free surface to be at the borderline between systems for which the solution band is limited by phase slip at the boundary (type I solutions) and those which have stable solutions over the entire bulk Eckhaus instability (type II solutions). Perhaps the transition from type I to type II solutions takes place by a gap opening near $k_c$, although a more likely course of events would involve $\lambda^\pm$ going to zero (i.e. a vanishing of the slopes of the straight lines in Fig.~\ref{fig:results_detail}). 

Perhaps a more likely explanation can be found in the finite length of the system. For the Eckhaus boundary of the finite system it is known that a gap opens up\cite{KZ85,HAC86} near $k_c$ which, for periodic boundary conditions,\cite{KZ85} has the width $\pm \delta k/k_c = \pm \pi/Lk_c$. Thus, any state with $-\pi/Lk_c < (k - k_c)/k_c < \pi/Lk_c$ is stable
all the way down to the neutral curve where its amplitude vanishes. For $L \simeq 50$ this gap in the Eckhaus boundary is shown in Fig.~\ref{fig:results_detail} by the two short dashed lines. One sees that the gap $\delta k^\pm/k_c$ derived from our 
data for $k_b^\pm(\epsilon)$ is of the same size; but as yet there is no theoretical prediction for finite-size effects on $k_b^\pm(\epsilon)$. 

Finally we comment on possible future experiments. Clearly the small-$\epsilon$, small-$(k - k_c)$ range needs to be examined by more detailed experiments to gain further insight into the nature of the gap discussed above. In part this can be done by varying the system length to see whether the gap is proportional to $L^{-1}$. Another interesting possibility for future work will be an attempt to alter the boundary conditions, thus effectively changing $\tilde \lambda$. It may be possible to do this by placing a less dense but perhaps more viscous fluid above the glycerol-water solution. Since the flow field of the Couette state is independent of the viscosity, one would not expect an Ekman vortex to be generated. However, the larger viscosity would lead to subcritical conditions in the added fluid when $\Omega = \Omega_c$ in the working fluid. The ensuing Couette state, located physically above the TVF state, should then reduce the TVF amplitude. This would correspond to $\tilde \lambda < 1$ and should yield a more narrow range of solutions (presumably larger $|\lambda^\pm|$) than those found in the present work.

\section{Acknowledgment}

We are grateful to Kapil Bajaj, Mark Urish, and Jun Liu for helpful discussions
about the experimental details. One of us (GA) is grateful to Lorenz Kramer for stimulating discussions about the theory, and to the Alexander von Humboldt Foundation for support and the University of Bayreuth for their hospitality while this paper was being written. This work was supported by the National Science
Foundation through Grant No. DMR94-19168. M.L. acknowledges support from the
Deutsche Agentur f\"ur Raumfahrtangelegenheiten (DARA).

\end{multicols}


\begin{references}
\bibitem{finite-size} Here we neglect the subtle effects on $k_E(\epsilon)$ associated with the finite size of the system which were discussed theoretically for periodic boundary conditions in Ref. \cite{KZ85} and which were studied experimentally with rigid non-rotating ends in Ref. \cite{HAC86}.
\end{references}
\end{document}